\documentclass[reprint,
superscriptaddress,
 amsmath,amssymb,
 aps,
pra,
]{revtex4-2}

\usepackage{graphicx}
\usepackage{dcolumn}
\usepackage{xcolor}
\usepackage{bm}
\usepackage{physics}
\usepackage{hyperref}
\usepackage{algorithm}
\usepackage{algpseudocode}
\hypersetup{
	colorlinks = true,
	urlcolor   = blue,
	linkcolor  = blue,
	citecolor  = red
}

\begin{document}

\title{SPIDERweb: a Neural Network approach to spectral phase interferometry }

\author{Ilaria Gianani}
\affiliation{Dipartimento di Scienze, Universit\`a degli Studi Roma Tre, Via della Vasca Navale 84, 00146 Rome, Italy}
\email{ilaria.gianani@uniroma3.it}

\author{Ian A. Walmsley}
\affiliation{QOLS, Department of Physics, Imperial College London, London SW7 2BW, UK}
\author{Marco Barbieri}
\affiliation{Dipartimento di Scienze, Universit\`a degli Studi Roma Tre, Via della Vasca Navale 84, 00146 Rome, Italy}
\affiliation{Istituto Nazionale di Ottica - CNR, Largo Enrico Fermi 6, 50125 Florence, Italy}

\begin{abstract}
Reliably characterised pulses are the starting point of any application of ultrafast techniques. Unfortunately, experimental constraints do not always allow optimising the characterisation conditions. This dictates the need for refined analysis methods. Here we show that neutral networks can provide a viable characterisation when applied to data from SPIDER. We have adopted a cascade of convolutional networks, addressing the multiparameter structure of the interferogram with a reasonable computing power. In particular, the necessity of precalibration is reduced, thus pointing towards the introduction of neural networks in more generic  arrangements.
\end{abstract}

\maketitle

The possibilities offered by the availability of ultrashort light pulses can can only be harnessed if an exhaustive, reliable, and accurate measurement of their temporal and spectral properties is available~\cite{Walmsley:09}. There exists a wide offer of techniques, that include Frequency-Resolved Optical Gating (FROG)\cite{Kane1993a}, Multiphoton Intrapulse Interference Phase Scan (MIIPS)\cite{Comin:16}, and Spectral Phase Interferometry for Direct Electric-field Reconstruction (SPIDER)\cite{Iaconis1998}, each further refined in variants tailored to tackle specific needs \cite{spider2,spider3,spider4,carspider1,Linden1998,Trebino2000}. These methods, with some adaptation, are also capable of operating at the quantum level~\cite{Maclean2019,brian,Gianani2019,PhysRevLett.121.083602, Karpinski21, PhysRevLett.129.123605}, thus further extending their range of operation. In this regime, however, data collection is typically more demanding than for detection at standard intensities. This motivates the need for improvements in the data processing that should be ideally resilient at lower signal-to-noise ratios. 
\begin{figure}[b!]
    \centering
    \includegraphics[width=\columnwidth]{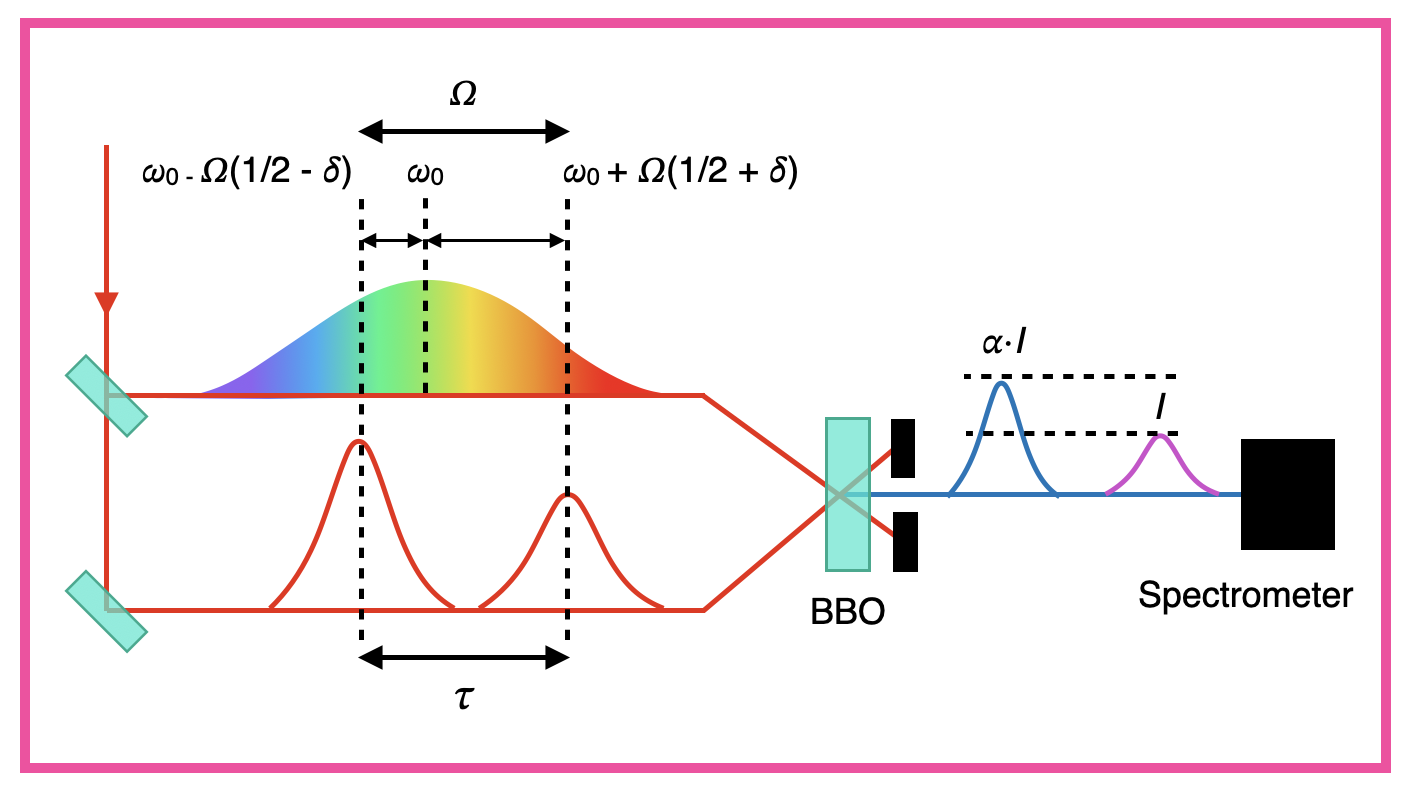}
    \caption{SPIDER scheme. Two replicas of the same pulse are upconverted on a nonlinear crystal together with a chirped ancillary pulse. The interferogram of the resulting signal is recorded with a spectrometer. Many parameters concur at defining its shape, notably the delay $\tau$ that also determines the shear $\Omega$; the relative amplitude $\alpha$; the division of the shear $\delta$;  }
    \label{fig:scheme}
\end{figure}

Among these, SPIDER is known for its robustness to noise and also for suffering from fewer ambiguities when compared to other techniques \cite{Anderson2000}. The price to pay is a more delicate experimental implementation, ending up requiring preliminary calibration steps which can be prone to errors and can complicate its use as a routine diagnostic tool. In this respect, neural networks (NNs) have emerged as a tool to navigate through datasets prone to the same problem: since they can be trained to recognised multiple parameters from the data, there exist the possibility of leveraging them in order to curtail the necessity of a preliminary calibration.  NNs have been introduced to ultrashort pulse characterisation~\cite{Zahavy:18,Kleinert:19,Stanfield2022,Kolesnichenko:23}, but not with the aim of reducing experimental requirements.

In this letter, we present a demonstration of this novel approach integrating NNs with SPIDER to remove the requirement for calibration. Specifically, we demonstrate how NNs do facilitate the spectral phase estimation in the absence of prior calibration, while allowing to relax the constraints on the shear.  We term this approach SPIDERweb. We employ our technique to experimentally evaluate the phase of a measured SPIDER interferogram and compare it with the standard algorithm. Our technique allows for the direct estimation of spectral phase coefficients, significantly simplifying the characterization process.

\begin{figure*}[t!]
    \centering
    \includegraphics[width=\textwidth]{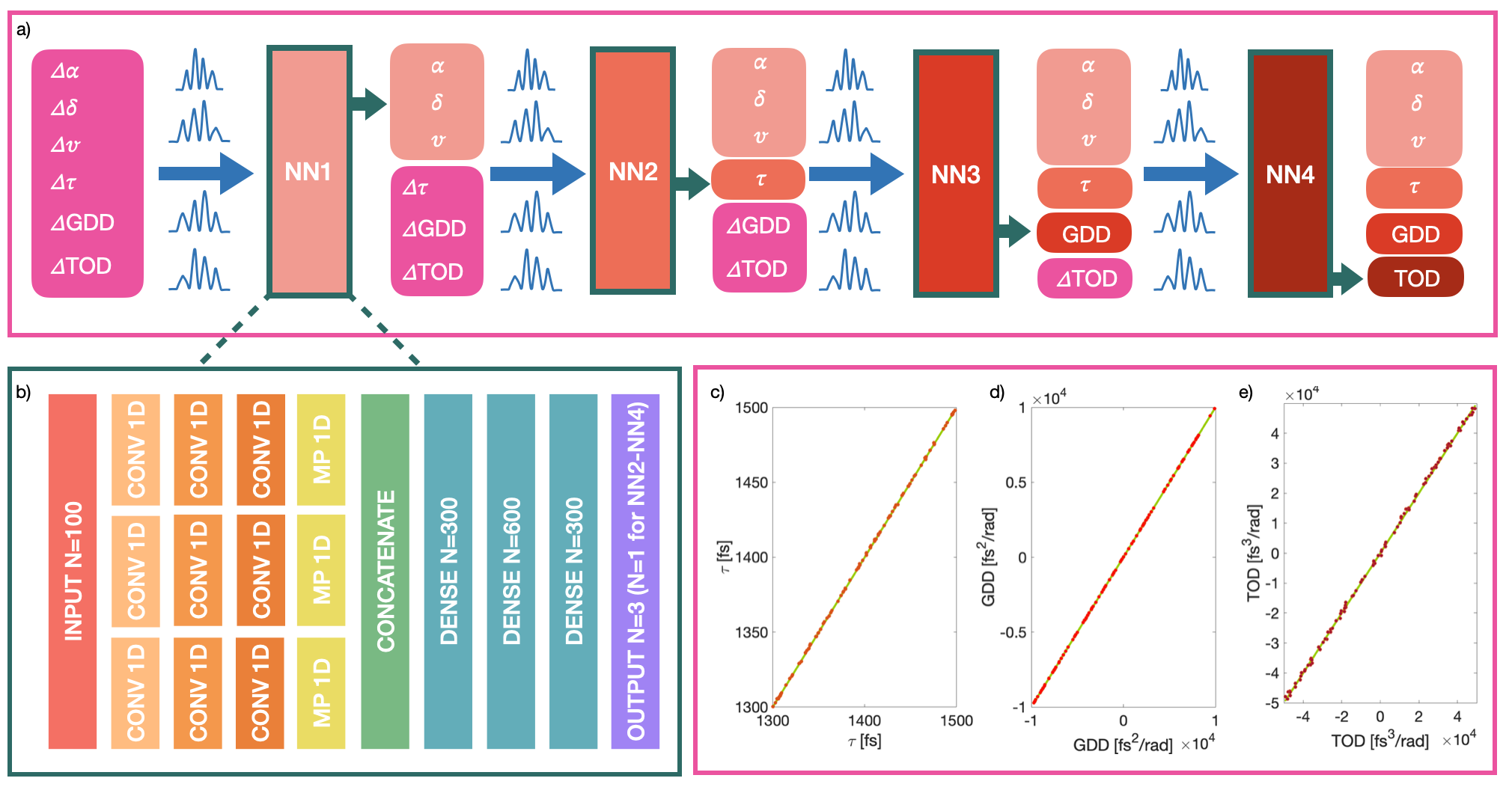}
    \caption{SPIDERweb. a) Scheme of the cascade NNs and the training procedure. b) Architecture of each individual NN. c) Test results for NN2, NN3, and NN4 in the retrieval of the time delay $\tau$ and the GDD and TOD coefficients respectively.} 
    \label{fig:NN}
\end{figure*}

In its essential features, a SPIDER measurement consists of the interference of two replicas of the same pulse $E(\omega)=A(\omega)\,e^{i\phi(\omega)}$ , delayed in time and sheared in frequency (Fig. \ref{fig:scheme}). In the most common implementation, a frequency shear is obtained by introducing an ancillary beam, originating from the same pulse, that has been stretched to be highly chirped. This ancilla is sent together with the two pulse replicas on a nonlinear crystal. Since the two replicas are delayed in time by $\tau$, they will upconvert with two different quasi-monochromatic slices from the chirped ancillary pulse. This will generate two signal pulses sheared in frequency by an amount $\Omega$ which is related to the time delay $\tau$ and to the group delay dispersion $GDD_a$ of the chirped pulse by $\Omega=\tau/GDD_a$. The interference between the two upconverted pulses is then recorded with a spectrometer, and reads: 
\begin{equation}
\begin{split}
    I_s(\omega) &= \vert E(\omega)\vert^2 + \vert E(\omega-\Omega)\vert^2 + \\
    &+2\vert E(\omega)\vert\vert E(\omega-\Omega)\vert\cos(\phi(\omega)-\phi(\omega-\Omega)+\omega\tau),
    \label{eq:spider}
\end{split}
\end{equation}
The phase $\phi(\omega)$ can be extracted via Fourier filtering by the Takeda algorithm \cite{Takeda:82}: if the delay $\tau$ is large enough, when Fourier transforming the pulse from the frequency to the time domain it is possible to isolate one of the two sidebands (AC) by amplitude filtering. Then, performing the inverse Fourier transform  on the filtered sideband allows to extract the phase difference: 
\begin{equation}
\theta(\omega) =\phi(\omega)-\phi(\omega- \Omega)+\omega\tau
\end{equation}
This is known as the SPIDER phase and can be used to construct the derivative of the phase $\phi(\omega)$ if the term $\omega\tau$ is removed. This is achieved by performing a calibration step consisting in recording an additional interferogram with $\Omega=0$, thus setting further requirements on the experimental apparatus.

In order to extract the spectral phase from $\theta(\omega)$ one can then proceed either via integration or via concatenation \cite{Dorrer2002}. Following for instance the concatenation route, once the phase for a given frequency $\omega_k$ is arbitrarily set, the phase on the next sampling point, $\omega_k + \Omega$ can be calculated as $\phi(\omega_k+\Omega)=\phi(\omega_k)+\theta(\omega_k+\Omega)$. The reconstructed phase will hence be sampled by the shear $\Omega$, which leads to strict constraints for the shear iteslf: in fact, it needs to be small enough to satisfy the Whittaker–Shannon sampling theorem ($\Omega \leq 2\pi/T$, where T is the temporal compact support of the pulse), but cannot be arbitrarily small as this would result in a higher sensitivity to noise and shot-to-shot instabilities \cite{Wyatt:11}. 
The retrieval process just described directly isolates the phase of the AC sideband to extract the phase, hence disregarding the information contained in the DC and in the amplitude of the AC components. While these do not carry information on the phase, they do contain information on the delay $\tau$ in the displacement of the AC sidebands, although its extraction with a fitting procedure would be prone to errors that would then negatively impact the phase estimation. For this reason in the standard phase retrieval algorithm the calibration step is preferred for evaluating $\tau$. Indeed, if one were to model the full interferogram, additional parameters would need to be included: 

\begin{itemize}
    \item in Eq. \ref{eq:spider} it is intrinsically assumed one of the pulses will be upconverted with its central frequency and will hence undergo no shear, while the other will be upconverted with the shear $\Omega$. The actual frequencies which will be involved in the upconverson however depend on the arrival time of the ancilla and the two replicas on the nonlinear crystal, so, in general, both pulses may be individually sheared by an arbitrary amount, albeit their overall frequency difference will be given by $\Omega$. We can then introduce an additional parameter $\delta$ so that one pulses may be sheared by $\Omega(1/2+\delta)$ and the other by $\Omega(1/2-\delta)$. 
    \item in Eq. \ref{eq:spider} we are also implying that the two copies of the pulse will be identical. Usually these are obtained by taking the front and back reflections of an etalon, thus while they will have the same spectral phase, they are bound to have different intensities: we can then introduce the additional parameter $\alpha$, so that defined one copy as $E(\omega)=A(\omega)\,e^{i\phi(\omega)}$, the other will be $E_{\alpha}(\omega)=\alpha E(\omega)$
    \item the interference fringes may have reduced visibility, so we need to introduce also a visibility parameter $\mathit{v}$.
\end{itemize}

 A more accurate description of the SPIDER interferogram is then: 
\begin{equation}
  \begin{split}
    &I_s(\omega) = \vert E(\omega+\Omega(1/2-\delta))\vert^2 + \vert \alpha E(\omega-\Omega(1/2+\delta))\vert^2 + \\
    &+2\alpha \mathit{v} \vert E(\omega+\Omega(1/2-\delta))\vert\vert E(\omega-\Omega(1/2+\delta))\vert\cdot\\
    &\cdot \cos(\phi(\omega+\Omega(1/2-\delta))-\phi(\omega-\Omega(1/2+\delta))+\omega\tau),
    \label{eq:spidercompleto}
\end{split}  
\end{equation}
The adoption of a parametric procedure to extract the values of these parameters would hardly be feasible. We now show that using a series of cascaded NNs instead, we can successfully extract the information on the phase from the SPIDER interferogram  without recurring to a calibration measurement. As customary, the spectral phase is described as a polynomial, here taken up to the third order: its reconstruction thus amounts to estimating the second (GDD) and third (TOD) order coefficients. Extensions to higher orders do not pose conceptual challenges. Since the extraction of the relevant phase is achieved by harnessing the information from the whole interferogram, we need to estimate, together with the phase and the time delay, also the additional parameters appearing in Eq. \ref{eq:spidercompleto} which will act as nuisance parameters in our protocol.  While this makes for a conspicuous number of parameters, we have found that adopting the cascaded NN structure as described in Fig. \ref{fig:NN}, allows us to keep the complexity of the network at a manageable level for it to be used as a routine diagnostic without becoming too computational resources-demanding. 
The training set is constructed by using the measured spectrum of the pulse in order to generate interferograms by assigning random values to all parameters involved. For each parameter $i$, we allow for an interval $\Delta_0i$. This set is adopted for the first network, NN1, extracting the values of the nuisance parameters $\alpha,\delta$ and $v$; once each network is trained, this is tested over newly generated interferograms, and these are used to evaluate the errors $\sigma_i$ , associated to the estimated parameters $i$. By this, we generate a new training set for the subsequent stage of the NN, with the estimated parameters restricted to the intervals $\Delta_1 i = [\bar i-3\sigma_i,\bar i+3\sigma_i ]$. In particular, NN2 is used to estimate the time delay $\tau$ within an uncertainty $\sigma_\tau$, and, in turn, these values and the previous inform the training set for NN3 (targeting the second-order dispersion GDD) and NN4 (targeting the third-order dispersion TOD). We here limit the estimation to the third order, as customary, but if higher orders need to be evaluated this can be extended adding further networks.
\begin{figure}[h!]
    \includegraphics[width=\columnwidth]{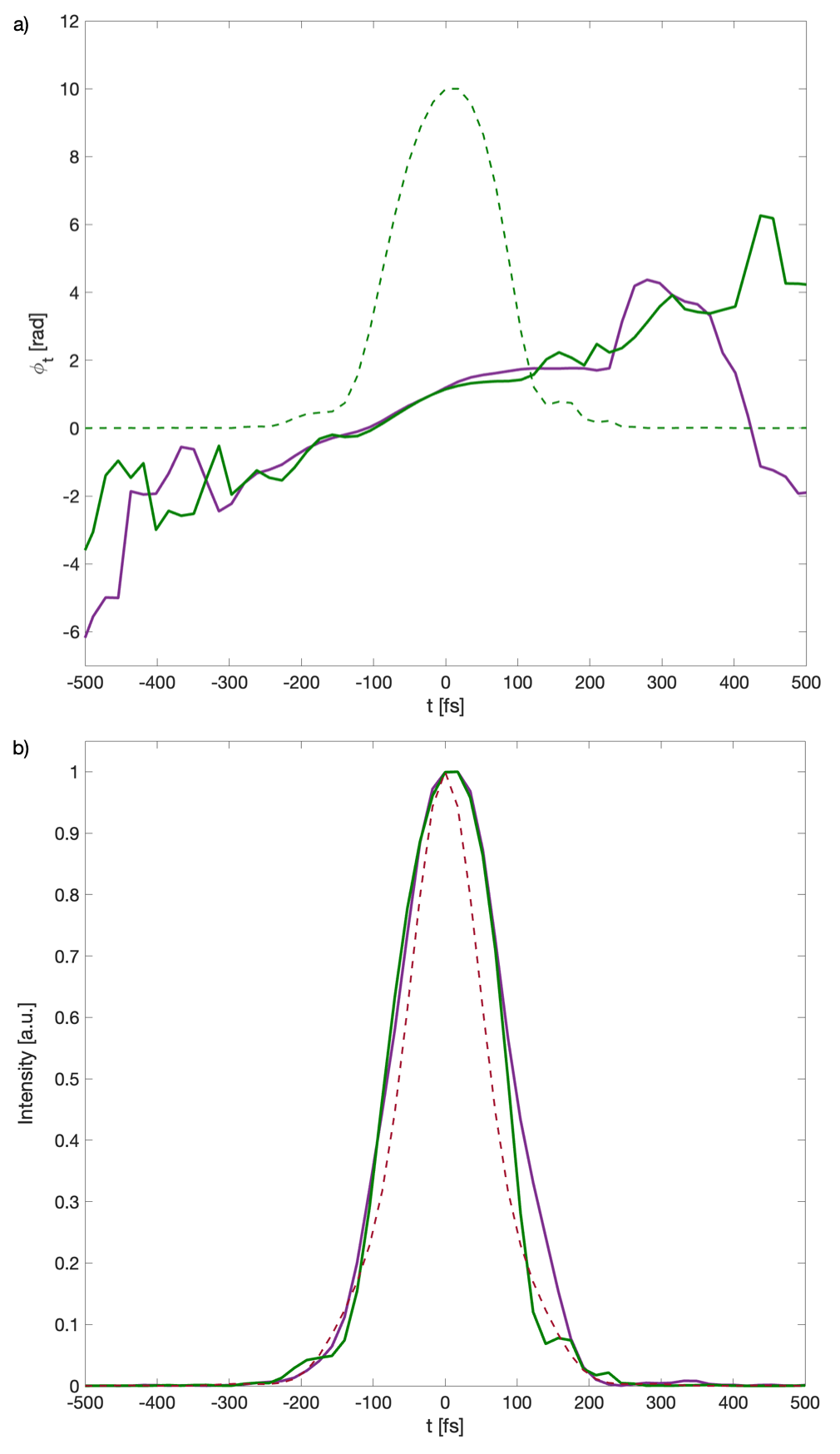}
    \caption{Reconstruction results. a) Temporal phase reconstructed using SPIDERweb (green) and the standard spider algorithm (purple). b) Temporal intensity reconstructed using SPIDERweb (green), the standard spider algorithm (purple), and the trasform-limited pulse obtained from the Fourier Transform of the measured spectra (dashed red).}
    \label{fig:results}
\end{figure}
The architecture of each $NN_i$ is the same, and is described in panel b) of Fig. \ref{fig:NN}: after the input layer, the network is split in three parallel convolutional 1D layers each operating with different kernel sizes (1,5, and 32). This allows the network to capture behaviours happening at different scales. The parallel structure is kept for three Convolutional 1D layers which vary for the number of filters applied ( 32, 16, and 8 for all three sub-networks) and a MaxPooling 1D level with pool size 2. These are then concatenated together and sent through three dense layers with N=300,600,300 neurons respectively. The output layer is then constituted by 3 neurons for NN1 and by 1 neuron for the other three networks. 

In order to test our approach on actual data, we used a home-built SPIDER device to measure a $\sim 200$-fs pulse centered around 807 nm. As a preliminary step, its spectrum is retrieved. Once this is performed, the reflection on an etalon creates the two test pulses, while the transmitted part is stretched by means of a double-grating arrangement. The beams are then focused on a $100$-$\mu$m Type-II BBO crystal by means of an off-axis parabolic mirror. This operation creates two frequency sheared copies, separated by $\Omega$. The value of $\Omega$ should normally be chosen judiciously, since the minimal duration of the pulse that can be revealed is of the order $\Omega^{-1}$, while the concatenation algorithm would demand small values of the shear. The upconverted signal is then collimated with a lens and filtered by means of a bandpass filter and two high-pass filters to ensure the removal of the residual red component and measured with a commercial spectrometer. This gives access to the SPIDER interferogram, and, by measuring the test pulses directly, to the calibration interferogram which will be required for the comparison with the standard algorithm. 

We create the training set for NN1 constituted by $N_{train}=3\cdot10^5$ interferograms generated starting from the measured spectrum and perform the training of NN1. We then use NN1 to estimate the values of the nuisance parameters using the measured SPIDER interferogram, and we proceed by training and running the networks as described above until all the experimental parameters are retrieved. For the training of each NN, we divide the generated datasets between training and validation with a 0.8/0.2 ratio and we use a batch size of $N_{batch}=500$, and run the training for $N_{epochs}=200$ epochs. Both the interferograms (features) and parameters (labels) are standardized and rescaled in each dataset to optimize the gradient descent. In panel c) of Fig. \ref{fig:NN} we show the results of the tests performed on NN2,3,4 for the estimation of $\tau$, GDD, and TOD respectively using in each instance $N_{test}=100$ newly generated interferograms. As explained above, these are used to evaluate the errors on the respective parameters.

The results of the estimation are shown in Fig. \ref{fig:results} where we report both the temporal phase and temporal intensity of the reconstructed pulse with the new method (green) against the estimation performed with the standard SPIDER algorithm (purple). 
The reconstructions appear very similar, as remarked by the root mean square percentage error (RMSPE) between the reconstructed temporal profiles of $1.09\%$. 

Our approach based on NNs allows to circumvent the need for the calibration step to derive $\tau$ from independent measurements - as in the standard SPIDER algorithm the stretcher $GDD_{\alpha}$ should be known in advance. The value of $\tau$ is actually extracted by means of the first stage in the cascaded NN, based on the same dataset employed for the full reconstruction. This structure is not necessary, in principle, and a single, larger newtwork could most probably deliver similar capabilities. However, we have found that this partitioned solution eases the requirement on the NN, making it possible to run the reconstruction on a computer with relatively modest computing power. Taking the path of a parametric approach for phase reconstruction also offers the advantage of curtailing the criticality of the choice of the shear $\Omega$. Since this does not dictate the resolution of the phase reconstruction, it does not conflict with the Whittaker-Shannon sampling conditions.

The main drawback of our approach is that a training procedure must be repeated for different spectra. This means that its applicability is rather suited as a diagnostic tools to real-time verification of the performance of light sources, in the same vein as the employ of optical spectrum analysers used to monitor lasers.

In conclusion, we have demonstrated the relevance of NN-based reconstruction for SPIDER. This resolves for the better some concerns that may be expressed due to the need for pre-calibration in the usual setting. On the contrary, the approach leveraging NNs aids easing these requirements to a large extent. This comes at the price of a more accurate modelling of the interferogram. Our results point towards the inclusion of such methods for the analysis of more refined arrangements, such as SEA-CAR-SPIDER~\cite{Witting:09}, that allow for multiple shears.

\section*{Acknowledgements} 
This work was supported by the European Commission (FET-OPEN-RIA STORMYTUNE, Grant Agreement No. 899587). IG acknowledges the support from MUR Dipartimento di Eccellenza 2023-2027.

\section*{Disclosures} The authors declare no conflicts of interest.

\section*{Data availability} Data underlying the results presented in this paper are not publicly available at this time but may be obtained from the authors upon reasonable request.

\bigskip


\bibliography{biblio}

\end{document}